\documentclass[a4paper]{article}
\usepackage[top=1.5in,bottom=1.5in,left=1.35in,right=1.35in]{geometry}
\usepackage{amsmath}

\usepackage{graphicx, color}
\graphicspath{{fig/}}

\title{Characterization of near-infrared to telecom frequency conversion in a rubidium-filled hollow-core photonic-crystal fiber}

\author{Jed A. Rowland$^1$, Chris Perrella$^{1,2,*}$, Rachel F. Offer$^1$,\\ Andre N. Luiten$^1$, Ben M. Sparkes$^{1,3}$ and Till J. Weinhold$^{3,4}$}

\date{
\begin{small} $^1$Institute for Photonics and Advanced Sensing (IPAS) and School of Physics, Chemistry and Earth Sciences, University of Adelaide, Adelaide, 5005 SA, Australia\\
$^2$Centre of Light for Life and School of Biological Sciences, University of Adelaide, Adelaide, South Australia 5005, Australia\\
$^3$Defence Science and Technology Group, Edinburgh SA 5111, Australia\\
$^4$Australian Research Council Centre of Excellence for Engineered Quantum Systems, School of Mathematics and Physics, The University of Queensland, Brisbane 4072, Australia\\
email: $^*$chris.perrella@adelaide.edu.au \end{small}}

\begin{document}
\maketitle

\begin{abstract}
We investigate near-infrared to telecommunications frequency conversion via a diamond four-wave mixing scheme in rubidium vapor contained within a hollow-core photonic-crystal fiber. 
The strong light-atom interaction in the fiber results in lower pump power requirements and higher conversion efficiency than can be achieved under equivalent conditions in a rubidium vapor cell. 
We also observe non-intuitive pump and signal frequency dependence of the four-wave mixing efficiency in the fiber due to the large nonlinearities present in the system.
These results indicate the potential for hollow-core fibers to provide a scalable solution to quantum information network infrastructure, with additional modelling required for a full understanding of the extreme atom-light interaction effects present.
\end{abstract}

\section{Introduction}
Quantum repeaters are a critical component for long-distance quantum information networks (QINs)~\cite{Kimble2008, Wei2022, Azuma2023}. 
To operate, quantum repeaters require quantum memories to synchronise probabilistic processes such as correlated pair production~\cite{Albrecht2014, Simon2010, Bussieres2013, Ma2020, Wei2022, Azuma2023}. 
Rubidium (Rb)-based memories have demonstrated many desirable qualities for this application, with room temperature operation allowing for storage efficiencies up to 87\,\%~\cite{Hosseini2011}, fidelities of 98\,\%~\cite{Hosseini2011}, gigahertz bandwidths~\cite{Finkelstein2018a}, and millisecond storage times ~\cite{Ebert2015}. 
However, these memories operate at near-infrared wavelengths, which is not conducive to low-loss propagation through current optical communications fibers.
Efficient conversion between near-infrared and telecommunication wavelengths is critical to be able to interface these Rb-based memories with low-loss fiber, thereby extending QIN ranges to significantly larger distances.

The conversion efficiency, $\eta$, is the primary metric for assessing the suitability of wavelength conversion between quantum memories in the near-infrared to telecommunication wavelengths.
$\eta$ quantifies the percentage of input photons that are converted to the new wavelength, with lower efficiencies reducing the information rates across QINs \cite{Jen2011}.
The amount of noise added during the conversion process is also a concern, as higher noise levels lead to a reduction in the signal-to-noise ratio (SNR) and fidelity of the process.
Another factor to consider is the infrastructure demand to scale the technology to a usefully sized network, e.g. optical power requirements.

There are many potential experimental schemes to convert near-IR photons to telecommunication wavelengths using non-linear processes, which conserve energy and momentum leading to strong quantum correlations.
For instance, solid-state approaches have included wave mixing that exploit second-order susceptibilities, $\chi^{2}$, within non-linear material such as periodically poled lithium niobate to produce sum and difference frequency generation.
Such frequency conversion processes have achieved; conversion of 616\,nm to 1552\,nm with efficiencies up to 62\% with 530\,mW of pump power~\cite{Maring2018}, conversion of 711\,nm to 1313\,nm with an efficiency of 64\% with 150\,mW of pump power~\cite{Zaske2012}, and conversion of 942\,nm to 1550\,nm with an efficiency of 56\% with 220\,mW of pump power~\cite{Morrison2021}. 

Ring resonators can also be used for frequency conversion via second- and third-order susceptibilities, the latter through four-wave mixing (FWM)  \cite{Becerra}. 
Wave-mixing in ring resonators has achieved high efficiency conversion due to their tight optical field confinement. 
For instance, wave-mixing in an aluminium nitride ring resonator has achieved 14\% conversion efficiency from 1530\,nm to 775\,nm with 28.3\,mW pump powers~\cite{Guo2016}, a ring resonator in diamond achieved 40\% conversion efficiency from 637\,nm to 1550\,nm with 50\,mW pump powers~\cite{Hausmann2014}, and a silicon ring resonator acheived 60\% conversion efficiency from 980\,nm to 1550\,nm with $<$60\,mW pump powers~\cite{Li2016}.
In all cases the pump powers required are impractical for scaling to large QINs~\cite{Maring2018, Zaske2012, Morrison2021}.

Another promising media in which to drive wave-mixing for near-infrared to telecommunications wavelength conversion is a Rb atomic vapor. 
Extensive research has been conducted on FWM in Rb, including work on double $\Lambda$~\cite{Yu2016, Liu2017} and diamond schemes~\cite{Whiting2018, Becerra, Willis2009a, Chaneliere2006, Franke-Arnold2010}. 
A key advantage of using a Rb vapor for frequency conversion is that it allows for the creation of a dual-purpose device that can function as both a quantum memory and a frequency converter.

The level structure of the $5S_{1/2}$, $5P_{1/2}$, $5P_{3/2}$, and $4D_{3/2}$ states of Rb provide a mechanism to convert 795\,nm photons to 1529\,nm (within the conventional fiber-optic C-band) via a FWM diamond scheme in the presence of 780\,nm and 1475\,nm pump beams~\cite{Gogyan2010, Willis2011, Radnaev2010}. Theoretical analysis predicts high fidelity conversion of qubits encoded in the photon number, path and polarization degrees of freedom, indicating the versatility of this scheme for a variety of quantum information processing protocols \cite{Tseng2024}.

\begin{figure}[t]
    \centering
    \includegraphics[width=.4\linewidth]{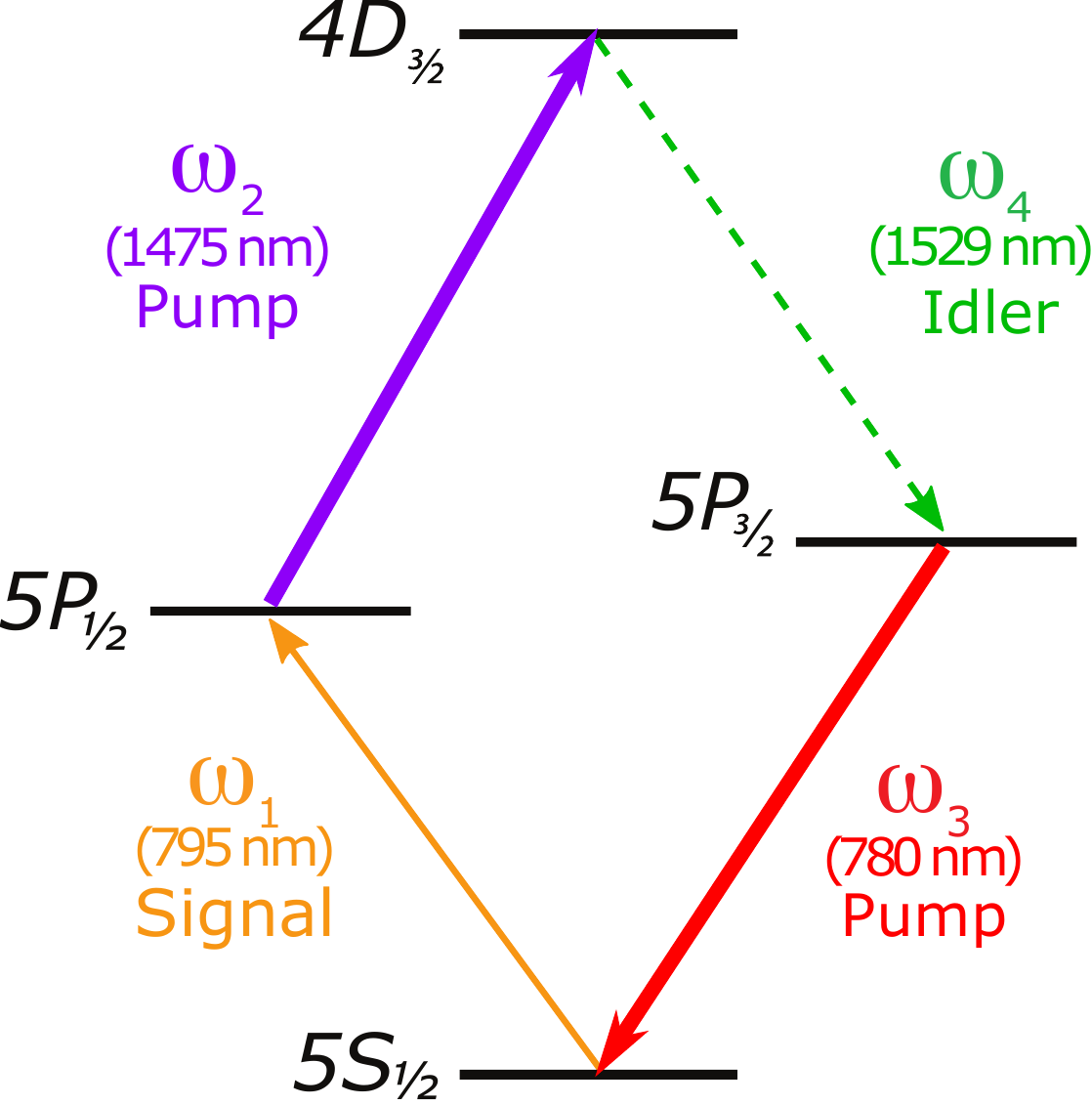}
    \caption{Rubidium atomic structure used for the conversion of a near-intrared 795\,nm signal to a telecom C-band 1529\,nm idler in the presence of strong pumps at 780\,nm and 1475\,nm via a diamond four-wave mixing scheme.}
    \label{fig:FWMscheme}
\end{figure}

This scheme is depicted in Fig.\ref{fig:FWMscheme}, with energy conservation and phase matching requiring~\cite{Boyd}: 
\begin{eqnarray}
    \omega_{4} & = &\omega_{1} + \omega_{2} - \omega_{3} \\
   \Vec{\kappa}_{4} & = &  \Vec{\kappa}_{1} + \Vec{\kappa}_{2} -\Vec{\kappa}_{3}
   \label{eqn:phasematching}
\end{eqnarray}
for maximising the efficiency of the frequency conversion. 
The phase mismatch is defined as
\begin{equation}
\Delta \Vec{\kappa}=\Vec{\kappa}_{1} + \Vec{\kappa}_{2} -(\Vec{\kappa}_{3}+\Vec{\kappa}_{4})
\label{eqn:phasemismatch}
\end{equation}
and has a dependence on the individual refractive indices, $n_i$, experienced by each wavelength in the FWM medium and also the relative angles, $\theta_i$ between each of the laser fields.

A previous experiment using the same diamond FWM scheme in a warm Rb vapor cell has shown a conversion efficiency of 6-8\% with pump Rabi frequencies of $2\pi \times 250$\,MHz and a co-propagating beam geometry (no angle phase matching) ~\cite{Piotrowicz, Kwolek2021}. 
Two experiments in cold atomic systems, with their much-increased atomic phase-space density, have shown improved efficiencies over warm atomic systems. 
Radnaev \textit{et al.} ~\cite{Radnaev2010}, demonstrated 54\% conversion efficiency between 795\,nm and 1367\,nm, and 30\% conversion efficiency between 795\,nm and 1529\,nm using cold $^{87}$Rb with pump powers of 4.3 and 13\,mW for the 780\,nm and 1475\,nm pumps respectively and optimal angle phase matching. 
Zhang \textit{et al.} \cite{ZhangFWM}, using a co-propagating beam geometry
in cold $^{85}$Rb, achieved a 32\% conversion efficiency between 795\,nm to 1529\,nm with pump powers of only 1.4\,mW and 1.87\,mW for the 780\,nm and 1475\,nm pumps respectively at an optical depth (OD) of 190. 
Detailed theoretical analysis of this diamond FWM scheme \cite{Tseng2024} 
predicts conversion efficiencies of greater than 80\% with co-propagating fields for ODs above 700, a value that is achievable in cold Rb \cite{Sparkes2013a, Halfmann2014}.  
However, cold atom systems require bulky and complex experimental apparatus including additional lasers for the cooling and trapping of the atoms, limiting their usefulness in a large-scale fiber-based QIN.

An alternative to cold-atoms for improving the interaction strength of an atomic medium is to use hollow-core photonic crystal fibers (HCPCFs).
The key advantage of HCPCFs is that they confine light within their hollow core that can be filled with Rb vapor. 
The small optical mode within their core allows generation of high intensity light, which can be maintained over the entire length of the fiber.
The arbitrary length of the vapor-filled HCPCF allows high ODs to be achieved~\cite{Poem2015, Sprague2013}, while high optical intensities allow efficient driving of non-linear atom-light interactions to be engineered~\cite{Perrella2018}. 
Together these properties should allow for increased efficiency of the FWM process while reducing the power requirements. 
However, the improved atom-light interaction strength that HCPCFs allow comes at the expense of fixing the  geometry of the beams to a strict co- or counter-propagating arrangement.
There will therefore be little ability to control the phase-matching condition shown in Eqns.\,\ref{eqn:phasematching} and \ref{eqn:phasemismatch} to ensure efficient FWM.
Even so, a diamond FWM scheme in a Rb-filled photonic bandgap fiber achieved 21\,\% conversion efficiency between a signal field at 776\,nm and idler field at 761\,nm with only 300$\mu$W of power in each pump~\cite{Donvalkar2014}.

In this article we focus on the FWM transition shown in Fig.\,\ref{fig:FWMscheme}, with a signal field at 795\,nm and idler field at 1529\,nm. 
The significant difference in wavelengths for this co-linear arrangement as compared to that in Ref.\,\cite{Donvalkar2014} will make phase matching more difficult.
We study the relationship between the detunings of the three input fields and between the powers of the two pump fields at 780\,nm and 1475\,nm on the FWM efficiency, and compare the performance in the HCPCF to a Rb vapor cell.
We show that the HCPCF offers higher conversion efficiencies over the cell for identical operating conditions. 
We observe significant differences in the spectral features of the FWM process between the cell and HCPCF operation due to the high intensities generated within the HCPCF, most notably a GHz-level shift in optimal pump and signal detuning parameters.
Exploitation of such features may make HCPCFs a platform suitable for use in QINs.

\section{Experiment}

\begin{figure}[b!]
    \centering
    \includegraphics[width=0.75\linewidth]{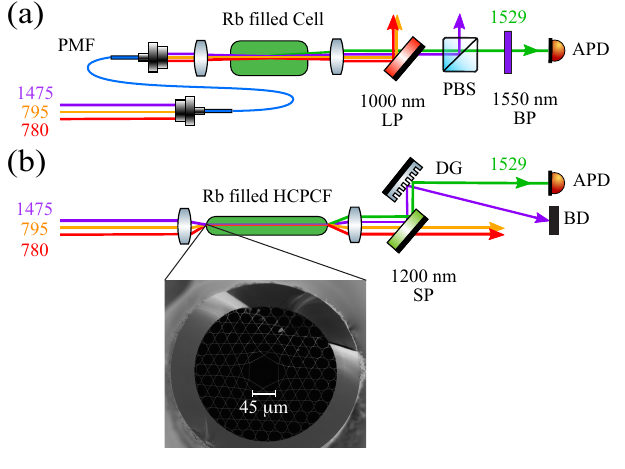}
    \caption{Schematic diagrams of the experimental setups utilizing (a) a vapor cell, and (b) a HCPCF in vacuum chamber. PMF: polarization maintaining fiber, optimised for guiding near-IR light, LP: long-pass dichroic mirror, SP: short-pass dichroic mirror, BP: $1550\,{\pm}\,25$\,nm band-pass filter, PBS: polarizing beam splitter, HCPCF: hollow-core photonic crystal fiber, DG: diffraction grating, APD: avalanche photo-diode, BD: beam dump. The inset shows a scanning electron microscope image of the $45$ $\mu$m core hollow-core fiber used.}
    \label{fig:experimentalsetup}
\end{figure}

The diamond FWM scheme we investigate for near-IR to telecommunication frequency conversion consists of a weak signal at 795\,nm (tuned close to the $5S_{1/2}\,{\rightarrow}\,5P_{1/2}$ $^{85}$Rb transition), and two pump fields at 1475\,nm (tuned close to the $5P_{1/2}\,{\rightarrow}\,4D_{3/2}$ transition) and 780\,nm (tuned close to the $5P_{3/2}\,{\rightarrow}\,5S_{1/2}$ transition). 
This arrangement produces an idler photon at 1529 nm (close to the $4D_{3/2}\,{\rightarrow}\,5P_{3/2}$ transition).

The 795\,nm signal field was generated by a tunable external cavity diode laser (ECDL) which was able to be frequency tuned over the full $5S_{1/2}\,{\rightarrow}\,5P_{1/2}$ manifold with its optical frequency being inferred from a saturated absorption spectroscopy (SAS) setup. 
The 1475\,nm pump field was produced by a cat-eye diode laser with a maximum output power of 67\,mW, which was frequency tunable over the $5P_{1/2}\,{\rightarrow}\,4D_{3/2}$ manifold, with an infrared optical wavemeter providing wavelength measurements. 
Finally, the 780\,nm pump field was generated by an ECDL with a maximum output power of 30\,mW, which was able to be frequency tuned over the $5S_{1/2}\,{\rightarrow}\,5P_{3/2}$ manifold, with a visible optical wavemeter providing wavelength measurements. 
The spectral line-width of each of these lasers was approximately 1\,MHz or less. 

Polarisation of the input fields is a critical factor in FWM phase matching. 
The highest efficiency for the FWM scheme was achieved when the 1475\,nm and 780\,nm pump lasers were horizontally polarized, while the 795\,nm signal laser was vertically polarized. 
This arrangement with orthogonal polarizations for the signal and pumps optimizes conservation of angular momentum~\cite{Chaneliere2006}, with the generated 1529\,nm idler, observed to be vertically polarised - the same as the signal.
The 780\,nm and 795\,nm fields were spatially overlapped via the transmission and reflection ports of an uncoated window, and were then combined with the 1475\,nm pump with a 1200\,nm short-pass (SP) dichroic mirror.

Once the lasers were spatially overlapped, they were directed towards either the HCPCF or a vapor cell.
The optical schematics for the cell and HCPCF are shown in Fig.~\ref{fig:experimentalsetup}(a) and (b) respectively.
The HCPCF naturally ensures that all three lasers are co-propagating and sharing a similar spatial mode. 
To allow for a direct comparison between the cell and HCPCF, the three lasers were delivered into the vapour cell via a 770-1100\,nm-guiding polarization maintaining fiber to ensure a co-linear geometry. 

In the 75 mm-long cell it was found that a $1/e^2$ intensity diameter of 100\,$\mu$m for the signal and 780\,nm pump, and 180\,$\mu$m for the 1475\,nm pump was optimal for FWM efficiency as a trade-off between increasing the peak intensity and reducing the Rayleigh range and thus interaction length.
The HCPCF used was a 30\,cm-long Kagome-style fiber with a 45\,$\mu$m core diameter. 
A cross-section of the HCPFC is shown in the inset of Fig.~\ref{fig:experimentalsetup}(b). 
The HCPCF guides light between 600-1700\,nm and supports an optical mode with a $1/e^2$ intensity diameter of $32.9\,{\pm}\,0.6$\,$\mu$m~\cite{Perrella2018}. 
While the HCPCF guides the 1475\,nm pump and 1529\,nm idler fields, these wavelengths experienced an additional 3\,dB of transmission loss over the full length of the fiber when compared to the 780\,nm pump and 795\,nm signal fields.

The HCPCF was held straight on a stainless-steel platform that sits within a vacuum chamber.
Both the HCPCF and vapor cell were filled with a natural isotope abundance rubidium vapor.
The vacuum chamber was heated to 80$^{\circ}$C to increase the Rb density, with the cell heated to a point which matched atomic density to that of the HCPCF. 

After passing through the cell or HCPCF, the three original wavelengths and the FWM generated 1529\,nm were spatially separated for detection. 
For the cell set-up, this was achieved with a long-pass dichroic mirror to remove the signal and 780\,nm pump, and then a polarizing beam-splitter and a $1550\,{\pm}\,25$\,nm band-pass filter to remove the 1475\,nm pump.
For the HCPCF set-up, this was achieved using a 1200\,nm short-pass dichroic to remove the signal and 780\,nm pump, and a diffration grating to separate the idler from the 1475\,nm pump.
A 1529\,nm laser, coupled through the HCPCF, was used to quantify the losses due to the dichroic and diffraction grating, allowing us to scale the measured idler power at the APD to the power directly out of the HCPCF and compare this to the power of the signal field transmitted through the HCPCF far from resonance.
This represents the internal efficiency of the FWM conversion process, without taking into account the estimated 3\,dB additional loss of the idler field inside the HCPCF.

\section{Results and Analysis}

\subsection{Frequency Detuning}
\begin{figure*}[t]
    \centering
    \includegraphics[width=1\textwidth]{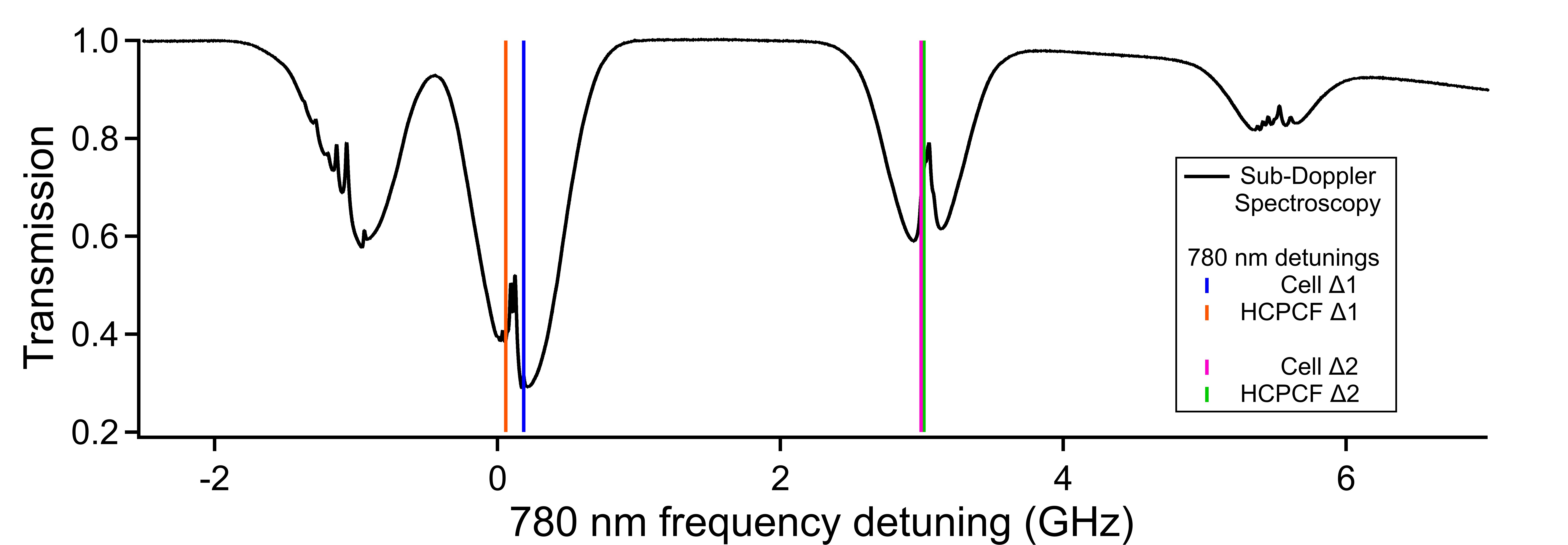}
    \caption{Saturated absorption spectroscopy of the 780\,nm $5S_{1/2}\,{\rightarrow}\,5P_{3/2}$ transition (black). Colored vertical lines show the positions of the 780\,nm pump detunings that produced optimal FWM efficiencies.
    Cell $\Delta1$ and HCPCF $\Delta2$ are +171 $\pm$20\,MHz and +53$\pm$\,20MHz detuned from the $5S_{1/2}(F\,{=}\,3)\,{\rightarrow}\,5P_{3/2}(F'\,{=}\,2)$ transition respectively (also the zero for the frequency axis). Cell $\Delta2$ and HCPCF $\Delta2$ are +16$\pm$20\,MHz and +38$\pm$20\,MHz detuned from the $5S_{1/2}(F\,{=}\,2)\,{\rightarrow}\,5P_{3/2}(F'\,{=}\,1)$ transition.} 
    \label{fig:satabs780}
\end{figure*}

\begin{figure*}[t!]
    \centering
    \includegraphics[width=1\textwidth]{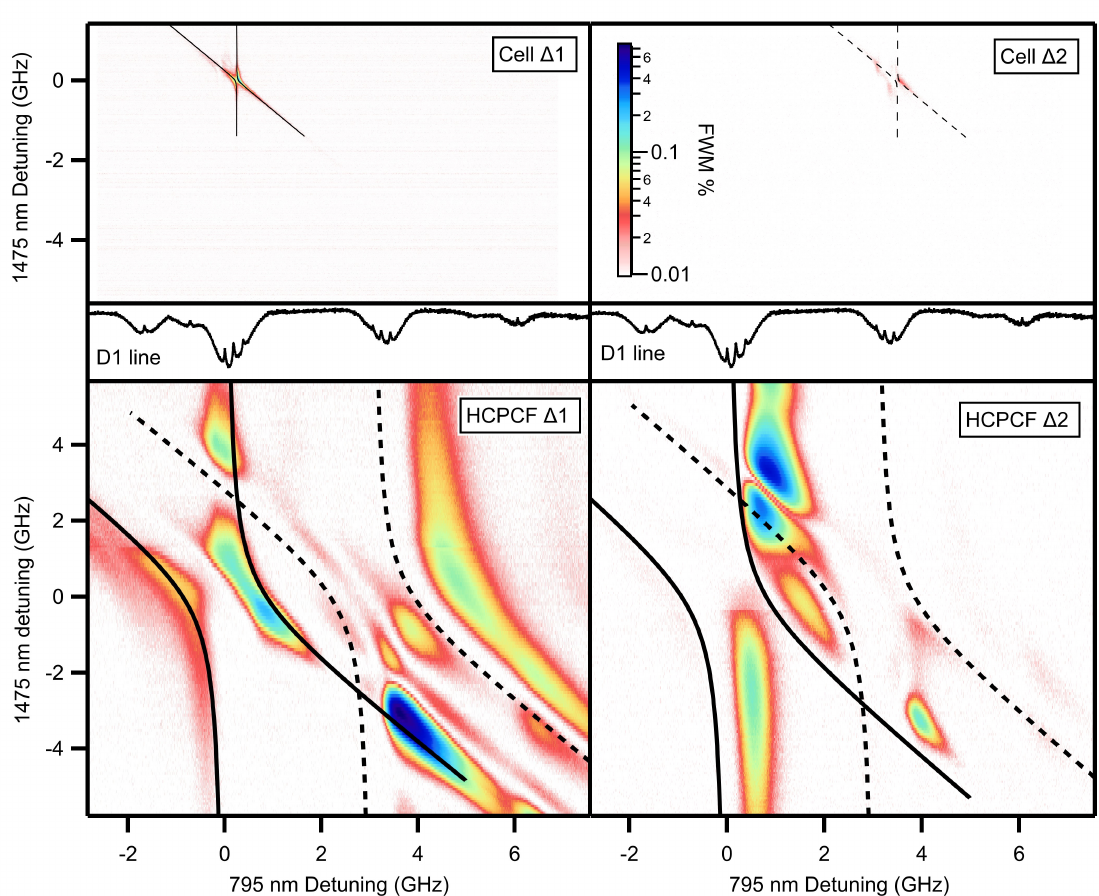}
    \caption{Two-dimensional color plots showing FWM efficiency (see legend) as a function of the detunings of the the 795\,nm signal laser and the 1475\,nm pump laser. 795\,nm detuning is taken from the $5S_{1/2}(F\,{=}\,3)\rightarrow5P_{1/2}(F'\,{=}\,2)$ transition and the 1475\,nm detuning is taken from the $5P_{1/2}(F'\,{=}\,2)\,{\rightarrow}\,4D_{3/2}(F''\,{=}\,1)$ transition. Note that the 1475\,nm scanning range is smaller for the cell data. 
    The corresponding 780\,nm pump detunings are shown in Fig.~\ref{fig:satabs780}. 
    Saturation absorption spectroscopy traces for the 795\,nm laser D1 line are included for reference. 
    Solid (dashed) black lines are avoided crossings of the form of Eq. \ref{eq:avoidedcrossing} and are not fitted to the data but overlaid as a visual guide (see text for details). For the Cell data $\Omega\,{=}\,75$\,MHz, while for the HCPCF data $\Omega\,{=}\,1.3$\,GHz. It is the frequency detunings of these areas that is depicted in Fig. \ref{fig:FWMschematics}.
    }
    \label{fig:2Dplots}
\end{figure*}

To allow comparison between the cell and HCPCF we set the temperature of each such that the atomic density was $9\,{\times}\,10^8$ atoms/cm$^3$.
This results in ODs of 13.5 and 54 in the cell and HCPCF, respectively.
To determine the optimal detunings for efficient FWM, the 1475\,nm pump was scanned over a frequency range of 10\,GHz centered on the $5P_{1/2}\,{\rightarrow}\,4D_{3/2}$ transition at a rate of 2\,mHz.
The 795\,nm signal was simultaneously scanned over the $5S_{1/2} (F\,{=}\,2,3)\,{\rightarrow}\,5P_{1/2} (F'\,{=}\,2,3)$ transition at a rate of 200\,Hz. 
The amount of idler field created as a function of the signal and 1475\,nm pump fields was captured on an oscilloscope at a rate of approximately 1\,Hz.
This means that there were approximately 500 traces for one continuously recorded 2-dimensional detuning map.
A smaller 1475\,nm pump detuning range was explored for the vapor cell, due to the lack of features present.
While the 1475\,nm pump and 795\,nm signal detunings were being scanned, the 780\,nm pump detuning was held constant near one of the two ground states as shown in Fig.\,\ref{fig:satabs780}.
These scans are shown in Fig.\,\ref{fig:2Dplots}.
We use the labels Cell $\Delta$1 and HCPCF $\Delta$1 for results from the cell and HCPCF respectively for when the 780\,nm pump laser frequency is tuned close to the $5S_{1/2} (F\,{=}\,3)\,{\rightarrow}\,5P_{3/2}$ transition.
We use the labels Cell $\Delta$2 and HCPCF $\Delta$2 for results from the cell and HCPCF respectively for when the 780\,nm pump laser detuning is tuned close to the $5S_{1/2} (F\,{=}\,2)\,{\rightarrow}\,5P_{3/2}$ transition.
Each of these configurations and results are described in detail below.

\begin{figure*}[t]
    \centering
    \includegraphics[width=1\textwidth]{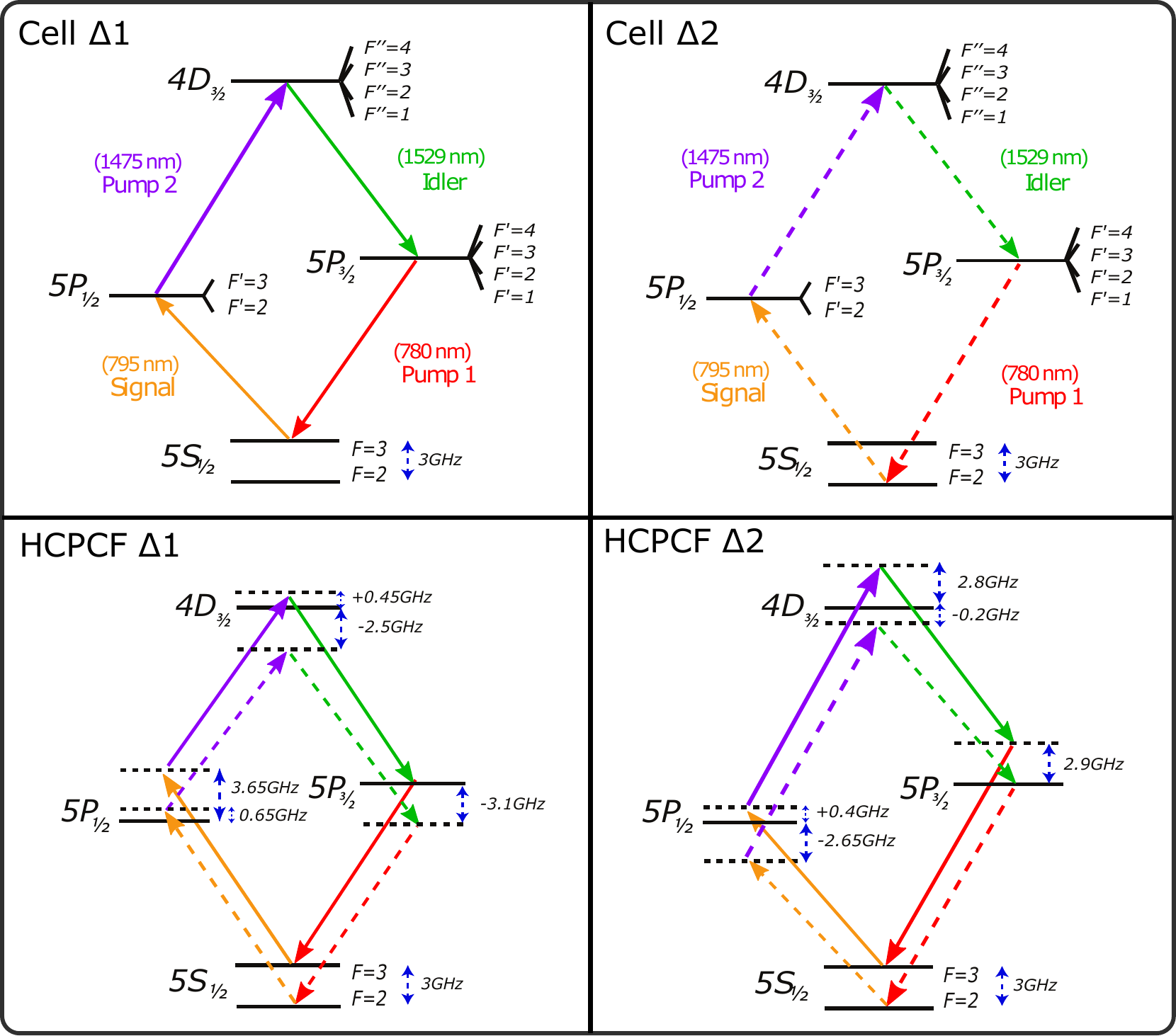}
    \caption{Level diagrams illustrating the laser frequency detunings that produce the largest FWM efficiency in each of the four configurations presented in Fig.~\ref{fig:2Dplots}. 
    Detunings presented in HCPCF $\Delta$1 and HCPCF $\Delta$2 diagrams refer to measured optical detunings from the weighted centre of the state manifold \cite{SteckRb85, Duspayev2023}.}
    \label{fig:FWMschematics}
\end{figure*}

Figure~\ref{fig:2Dplots} (Cell $\Delta$1) shows the vapor cell FWM conversion efficiency with the 780\,nm pump placed approximately on resonance with the $5S_{1/2}(F\,{=}\,3)\,{\rightarrow}\,5P_{3/2} (F'\,{=}\,4)$ transition (orange line in Fig.\,\ref{fig:satabs780}). 
The 795\,nm signal power was set to 100\,$\mu$W, below the single-photon saturation threshold for the cell.
The 1475\,nm pump was set to 5.5\,mW and the 780\,nm pump was set to 1.5\,mW. 
These parameters produced a maximum FWM efficiency of 0.2\% with the signal resonant with the $5S_{1/2} (F\,{=}\,3)\,{\rightarrow}\,5P_{1/2}(F'\,{=}\,2,3)$ transition and the 1475\,nm pump resonant with the $5P_{1/2}\,{\rightarrow}\,4D_{3/2}$ transition, as shown in Fig.~\ref{fig:FWMschematics}. 
The dominant feature corresponds to two photon resonance with the $5S_{1/2} (F\,{=}\,3)\,{\rightarrow}\,4D_{3/2}$ transition, with a strong avoided crossing observed when the 795nm signal is single photon resonant with the $5S_{1/2} (F\,{=}\,3)\,{\rightarrow}\,5P_{1/2}(F'\,{=}\,3)$ transition. 
This behaviour is commonly seen in similar diamond FWM schemes in Rb~\cite{Whiting2018, Becerra}.
A second, much weaker, avoided crossing due to the $F'{=}2$ hyperfine state of the $5P_{1/2}$ level can also be seen.

The solid black lines indicate the two-photon transition avoided crossings of the form \cite{Culver2016, Rapol2002}
\begin{equation}    
    \Delta_{1475} = \frac{1}{2}\left(-\Delta_{795}\pm \sqrt{\Delta_{795}^2+\Omega^2}\right),
    \label{eq:avoidedcrossing}
\end{equation}
where $\Delta_i$ is the detuning of the laser of wavelength $i$ and $\Omega$ is the Rabi frequency of the two-photon transition.
The black line in Fig\,\ref{fig:2Dplots} (Cell $\Delta$1) shows an avoided crossing using the equation above with a Rabi frequency of $\Omega\,{=}\,75$\,MHz.
Figure~\ref{fig:2Dplots} (HCPCF $\Delta$1) shows the result of repeating the  detuning scan from Cell $\Delta$1 but in the HCPCF. The 780\,nm pump was red-detuned by 120\,MHz from its position for Cell $\Delta$1 data (blue line in Fig.~\ref{fig:satabs780}) to optimize the FWM efficiency for this arrangement. 
For the data shown in Fig.\,\ref{fig:2Dplots} (HCPCF $\Delta$1), the signal power was set to 150\,$\mu$W, the 1475\,nm pump was set to 1.43\,mW, and the 780\,nm pump was set to 720\,$\mu$W to optimize the FWM efficiency.
All these powers were well above single-photon saturation for their respective transitions within the HCPCF which is on the order of $0.1\,\mu$W~\cite{Perrella2012}.
These optical powers are also comparable to, or greater than, two-photon saturation powers previously observed within the HCPCF~\cite{Perrella2013}.
It is immediately obvious that there are many additional features present in the HCPCF data to that from the cell.
To understand this better, we include avoided crossings from both ground states (i.e., setting $\Delta_{795}=0$\,GHz (solid line) and $\Delta_{795}=3$\,GHz (dashed line) for $F = 3$ and 2 respectively using Eq.~\ref{eq:avoidedcrossing} with $\Omega = 1.3$\,GHz. 
The ratio of Rabi frequencies between the HCPCF and cell data was approximately 16, which is consistent with the ratio of electric fields between the 780\,nm pumps setups which was approximately 12.
The maximum FWM efficiency of 0.75\% seen here is a factor of 3.5 higher than that from the cell.
All four wavelengths were measured at this point of maximum efficiency.
The difference between the sum of signal and 1475\,nm pump, and the idler and the 780\,nm pump was $-26\pm 72$\,MHz, within the combined resolution limit of the wavemeters used to make the measurements, indicating the closed-loop nature of the FWM process being driven.
These measured frequencies are consistent with two potential FWM arrangements, one resonant with each ground state, as shown in Fig.~\ref{fig:FWMschematics} (HCPCF $\Delta$1).
The unpumped $^{85}$Rb-filled HCPCF, with approximately equal ground state population, would allow both these FWM arrangements to be driven simultaneously. 
This behavior is similar to that observed previously in Ref.\,\cite{Franke-Arnold2010}.

In contrast to the cell data, in the HCPCF strong FWM is observed when the 795\,nm signal is resonant with both hyperfine ground states.  
It is also interesting to note that where the avoided crossings from the two ground states overlap, we see a reduction in FWM conversion efficiency. 
This appears to indicate destructive interference occuring between the two possible FWM arrangements.
The other features seen in Fig.\,\ref{fig:2Dplots} (HCPCF $\Delta 1)$ are likely due to the interaction between the various hyperfine levels of the driven transitions, as well as depletion of the signal and idler fields within the highly absorptive HCPCF system when they are close to one-photon resonance.

Figure~\ref{fig:2Dplots} (Cell $\Delta$2) shows the 2D FWM conversion efficiency map with the 780\,nm pump placed 16\,MHz blue-detuned from the $^{85}$Rb $5S_{1/2} (F\,{=}\,2)\,{\rightarrow}\,5P_{3/2} (F'\,{=}\,1)$ transition (purple line in Fig.~\ref{fig:satabs780}). 
The corresponding FWM arrangement for the maximum conversion efficiency is shown in Fig.~\ref{fig:FWMschematics} (Cell $\Delta$2).
The laser powers used were identical to those for Cell $\Delta$1.
Using the same Rabi frequency in Cell $\Delta$1, we see similar match with the avoided crossings, but with a lower maximum FWM conversion efficiency of 0.014\%.
 
Figure~\ref{fig:2Dplots} (HCPCF $\Delta$2) shows the result of repeating the detuning scan from Cell $\Delta$2 but in the HCPCF.
The 780\,nm pump was 16\,MHz further blue-detuned from its position for Cell $\Delta$2 data (green line in Fig.\,\ref{fig:satabs780}) and the laser powers were identical to that of HCPCF $\Delta$1.
We again include avoided crossings to the data, using the same Rabi frequency as HCPCF $\Delta$1, and  see the same key characteristics.
Namely that the overlap of the two avoided crossings leads to a reduction in FWM conversion efficiency.
The maximum FWM conversion efficiency in this arrangement of 0.47\% again appears adjacent to the overlap of the two avoided crossings.
The sum of the frequencies was $-85\pm 72$\,MHz, just outside the expected error of the wavemeters used.
This may be due to slow frequency drifts of the unstabilized lasers during the measurement.
Figure~\ref{fig:FWMschematics} (HCPCF $\Delta$2) illustrates the two potential FWM arrangements responsible for this maximum efficiency.

\subsection{Power Dependence}
\begin{figure}[t!]
    \centering
    \includegraphics[width=0.75\linewidth]{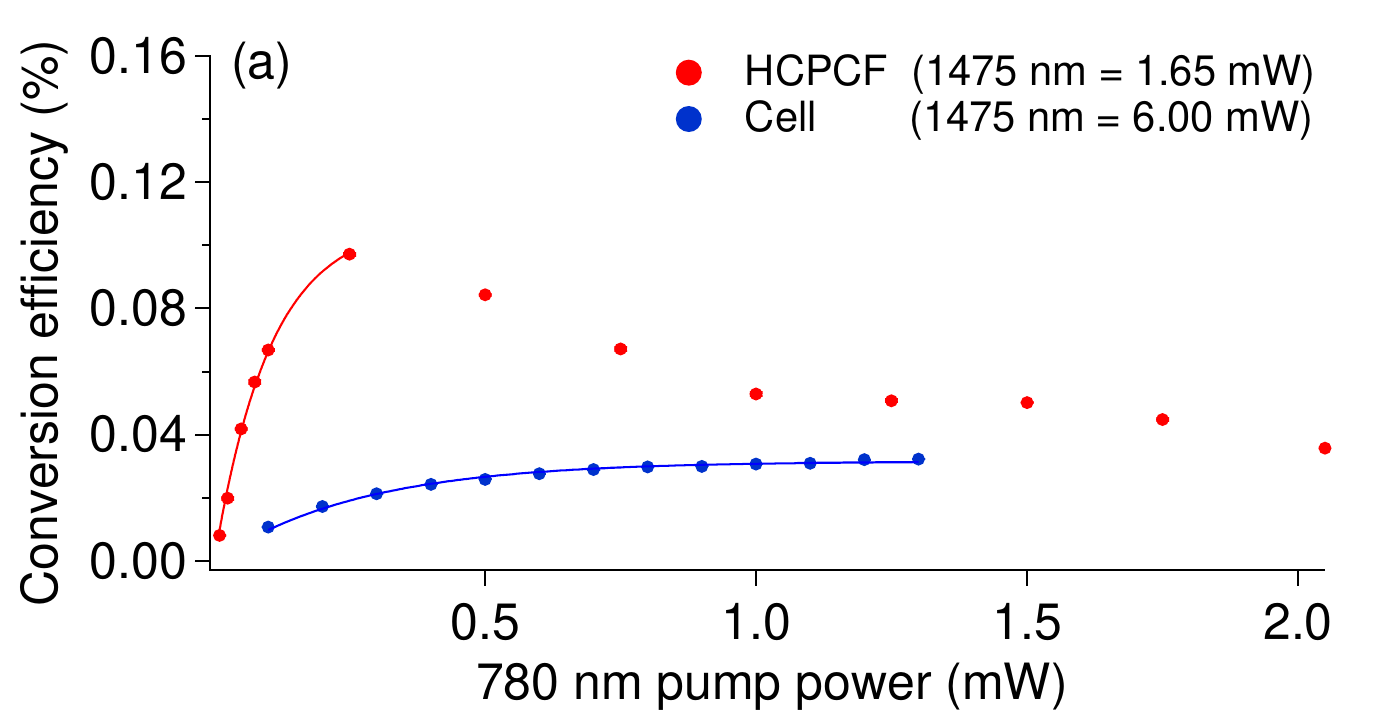}  \includegraphics[width=0.75\linewidth]{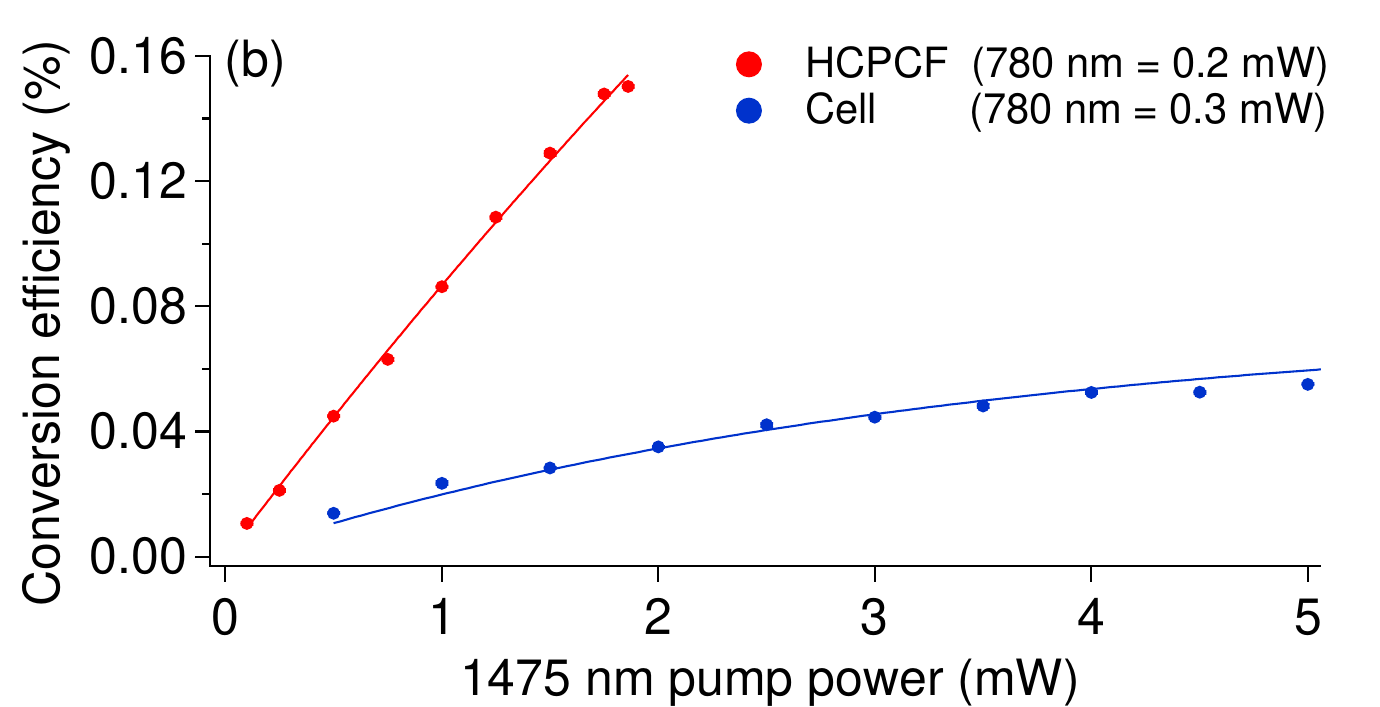}
    \caption{Four-wave mixing efficiency as a function of (a) 780\,nm pump power and (b) 1475\,nm pump power for both the hollow-core photonic crystal fiber (HCPCF) and vapor cell (Cell).}
    \label{fig:powers}
\end{figure}

To monitor the power dependence of the FWM process, we set the temperature of each such that the atomic density was $3\,{\times}\,10^8$ atoms/cm$^3$.
This results in ODs of 4.5 and 18 in the cell and HCPCF, respectively.
Figure~\ref{fig:powers} shows the FWM conversion efficiency dependence upon the 780\,nm and 1475\,nm pump laser powers for both the HCPCF and cell.
The frequency detunings of the lasers were set to the most efficient arrangement determined from the frequency detuning maps shown in Fig.\,\ref{fig:2Dplots}. 
In the HCPCF, a 795\,nm power of 200\,$\mu$W was used, while in the vapor cell a 795\,nm power of 100\,$\mu$W was used to ensure measurable idler detection signals even at low pump powers.  

The dependence of the FWM efficiency in the HCPCF and cell upon the 780\,nm pump field power is presented in Fig.~\ref{fig:powers}(a).
Both the HCPCF and cell efficiencies initially increase with 780\,nm optical power.
For the cell, the efficiency flattens to a near constant value of 0.031\% with a 1475\,nm pump power of 6.0\,mW, while for the HCPCF we see a decrease in efficiency once the 780\,nm pump power passed 250\,$\mu$W, potentially due to a number of effects such as power broadening or Autler-Townes splitting~\cite{Whiting2018,  DeSouza2022}, with a maximum efficiency of 0.097\% with a 1475\,nm pump power of 1.65\,mW and 780\,nm power of 250\,$\mu$W.
In both cases, the 1475\,nm pump power was taken to be the maximum available.
These trends are fitted with a saturation equation of the form~\cite{Radnaev2010}:
\begin{equation}    
    \eta(P) = \eta^\textrm{max}\left[1-\exp(-P/P_\textrm{sat})\right],
\label{efficiency}
\end{equation}
where $\eta^\textrm{max}$ is maximum efficiency, $P$ is the pump power and $P_\textrm{sat}$ is the pump saturation power. 

Figure~\ref{fig:powers}(b) shows the conversion efficiency dependence upon the 1475\,nm pump beam power for the HCPCF and vapor cell, with Eqn.~\ref{efficiency}  used for the power depencence of the 1475\,nm pump.
For the HCPCF, we placed the 780\,nm pump power at the point of maximum measured efficiency from Fig.~\ref{fig:powers}(a), and used a similar value for the cell data.
In both cases, the maximum efficiency occurs with the maximum 1475\,nm pump power available.
For the HCPCF this corresponds to a maximum efficiency of 0.15\% with 1.86\,mW of 1475\,nm pump power.
For the cell, this corresponds to a maximum efficiency of 0.055\% with 5.0\,mW of 1475\,nm pump power.

\section{Discussion}
The results presented above highlight the utility of HCPCFs for fundamental studies of atom-light interactions in the strongly-coupled regime.
The enhanced atom-light interaction strength mediated by the Rb-filled HCPCF system has led to  an approximately four-fold increase in the FWM conversion efficiency for a constrained co-linear geometry compared to a cell with the same atomic density, and provides a new environment within which to study FWM.
A possible explanation for the different behavior observed between the cell and the HCPCF may include the combination of single-photon absorption close to individual pump, signal and idler resonances, as well as destructive interference between FWM arrangements originating from the two ground states as shown in the bottom half of Fig.\,\ref{fig:FWMschematics}.
To confirm and understand this behavior a more detailed modelling effort describing more atomic levels and transitions would be required, which is beyond the scope of this work. 
Alongside the observation of the novel spectral features we confirm the significant reduction in pump power requirements for this frequency conversion process in HCPCFs due to their small mode areas. 
This identifies HCPCFs as being well suited for efficient single-photon-level signal operations, as envisaged for a QIN operating on classical telecommunication fiber networks with embedded atomic memories.

The use of HCPCF naturally restricts beams to be co-linear, removing the ability to minimise the phase mismatch via beam angles. 
Our work provides a preliminary investigation into the complexity of phase-matching within a co-linear geometry, where we only utilized the detuning of the signal and pump beams to optimise FWM efficiency. 
If we relax the co-linear geometry requirement, it has been shown that the angles which optimise phase-matching are between 1-2$^\circ$ \cite{Chaneliere2006}.
In our cell, with a small angle between the beam, we can see FWM efficiencies up to approximately 4\%, compared to the 0.2\% reported above.
However, considering the large FWM parameter space to be explored, if modelling were to be performed it could potentially lead to the identification of regimes of greater FWM conversion efficiency than those observed here.

There are other mechanisms by which the efficiency could also be increased within the HCPCF system. 
For instance, previous theoretical work on diamond FWM schemes in Rb has predicted and verified that larger ODs result in larger efficiency~\cite{Jen2011, Whiting2018, ZhangFWM, Tseng2024}. 
Increasing the OD of the HCPCF system can be achieved through direct heating of the fiber to create a persistently high OD.
The largest OD utilized in this work was ${\sim}\,54$ whereas an OD of 120 has been previously observed in the same HCPCF system. 
Using the model presented in \cite{ZhangFWM} we estimate that this increase in OD would result in an increase in efficiency to ~5-10\%. 
The OD can also be increased temporarily through pulsed techniques such as Light-Induced Atomic Desorption (LIAD), which have been shown to achieve ODs $>$ 2000 for hours at a time, and over $10^4$ for much shorter periods ~\cite{Poem2015, Sprague2013}. At these OD levels, while the optimal detuning parameters may change, the conversion efficiency should reach over 85\% \cite{Tseng2024}.

Apart from simply increasing the OD via the number density, control of the atomic states can increase the light-atom coupling strength.
Optical pumping of $^{87}$Rb in HCPCF has recently been demonstrated ~\cite{Krehlik2022} and can be used to increase the number of atoms in a single ground state for FWM~\cite{Chaneliere2006, Akulshin2011}. 
This can lead to substantial increases in efficiency up to an order of magnitude~\cite{Akulshin2011} while simultaneously reducing the complexity of pathways through magnetic sub-states and therefore simplify the detuning structure observed. 
Electromagnetically Induced Transparency (EIT) has also been regularly used to improve efficiency in diamond FWM schemes~\cite{Willis2009a, Wen2014, Deng2002} by reducing linear absorption of driving lasers and increasing the non-linear susceptibility. 
Finally, if we were to use an isotopically pure $^{87}$Rb vapor, then the complexity of available transitions could be further reduced, removing unwanted single photon absorption and other potentially competing processes in $^{85}$Rb~\cite{Willis2009a, Chaneliere2006}. 

Another mechanism to increase the FWM efficiency is through careful design of the HCPCF itself.
The HCPCF used here was not ideal for guidance at the very different wavelengths required for this FWM scheme.
This led to the HCPCF attenuating the 1529\,nm idler by a factor of 3\,dB more than the 795\,nm signal.
Recent developments in HCPCF design and manufacture has enabled low-loss guidance over much broader wavelength ranges than we use here~\cite{Nasti2022}, that would significantly improve the FWM efficiency. 
Furthermore, utilising a HCPCF with a smaller diameter core could reduce required pump powers even further~\cite{Donvalkar2014}. 
Finally, birefringent HCPCFs have been developed which could enable improved phase-matching even in a co-linear arrangement~\cite{Hosen:22}. 
The ability to join HCPCFs to single-mode fibers (SMFs) has recently become a viable option with achieved coupling efficiencies up to 84.5\%~\cite{Fan2017} and 97\%~\cite{Suslov2021}. 
This  will improve the end-to-end efficiency of the HCPCF FWM system, as well as increase the ease of integration of the HCPCF FWM system into existing fiber based networks~\cite{Sprague2014, Londero2009, Krehlik2022}. 

\section{Conclusion}

We have studied a diamond FWM scheme in a Rb-filled HCPCF to determine if HCPCFs provides a promising platform for quantum frequency conversion. 
We investigated the detuning and pump power dependence of co-linear FWM in a vapor cell and then compared this to the HCPCF. 
We showed that the HCPCF can produce a nearly four-fold higher conversion efficiency (0.75\%) using two-to-four-fold lower pump powers. 
We observed significant differences in detuning behaviour between the cell and the HCPCF, made possible by the high intensities maintained in the fiber, that utilizes both ground states and therefore 100\% of the $^{85}$Rb atoms. 
These results indicate that, with a carefully selected HCPCF and experimental improvements, HCPCFs can provide a scalable technology for use in QINs.\\

\subsubsection*{Funding}
B.M.S. acknowledges support from an Australian Research Council (ARC) Discovery Early Career Researcher Award (Grant No. DE170100752). 
J.R. acknowledges the support he has received for his research through the provision of an Australian Government Research Training Program Scholarship.
This research was supported by the Australian Research Council Centre of Excellence for Engineered Quantum Systems (EQUS, CE170100009). 

\subsubsection*{Acknowledgment}
The authors thank A. Black and M. Bashkansky for useful discussions and insight.
This work was supported by the Commonwealth of Australia as represented by the Defence Science and Technology Group of the Department for Defence.

\subsubsection*{Disclosures}
The authors declare no conflicts of interest.

\subsubsection*{Data Availability Statement}
Data underlying the results presented in this paper are not publicly available at this time but may be obtained from the authors upon reasonable request.

\bibliographystyle{unsrt}
\bibliography{Bibliography}

\begin{thebibliography}{10}

\bibitem{Kimble2008}
H.~J. Kimble.
\newblock {The quantum internet}.
\newblock {\em Nature}, 453(7198):1023, 2008.

\bibitem{Wei2022}
Shi-Hai Wei, Bo~Jing, Xue-Ying Zhang, Jin-Yu Liao, Chen-Zhi Yuan, Bo-Yu Fan, Chen Lyu, Dian-Li Zhou, You Wang, Guang-Wei Deng, Hai-Zhi Song, Daniel Oblak, Guang-Can Guo, Qiang Zhou, S.-H Wei, B~Jing, X.-Y Zhang, J.-Y Liao, C.-Z Yuan, B.-Y Fan, C~Lyu, D.-L Zhou, Y~Wang, G.-W Deng, H.-Z Song, G.-C Guo, Q~Zhou, and D~Oblak.
\newblock {Towards Real-World Quantum Networks: A Review}.
\newblock {\em Laser {\&} Photonics Reviews}, 16(3):2100219, 2022.

\bibitem{Azuma2023}
Koji Azuma, Sophia~E. Economou, David Elkouss, Paul Hilaire, Liang Jiang, Hoi~Kwong Lo, and Ilan Tzitrin.
\newblock {Quantum repeaters: From quantum networks to the quantum internet}.
\newblock {\em Reviews of Modern Physics}, 95(4):045006, 2023.

\bibitem{Albrecht2014}
Boris Albrecht, Pau Farrera, Xavier Fernandez-Gonzalvo, Matteo Cristiani, and Hugues de~Riedmatten.
\newblock {A waveguide frequency converter connecting rubidium-based quantum memories to the telecom C-band}.
\newblock {\em Nature Communications}, 5(1):1--6, 2014.

\bibitem{Simon2010}
C.~Simon, M.~Afzelius, J.~Appel, A.~{Boyer De La Giroday}, S.~J. Dewhurst, N.~Gisin, C.~Y. Hu, F.~Jelezko, S.~Kr{\"{o}}ll, J.~H. M{\"{u}}ller, J.~Nunn, E.~S. Polzik, J.~G. Rarity, H.~{De Riedmatten}, W.~Rosenfeld, A.~J. Shields, N.~Sk{\"{o}}ld, R.~M. Stevenson, R.~Thew, I.~A. Walmsley, M.~C. Weber, H.~Weinfurter, J.~Wrachtrup, and R.~J. Young.
\newblock {Quantum memories}.
\newblock {\em The European Physical Journal D}, 58(1):1--22, 2010.

\bibitem{Bussieres2013}
F{\'{e}}lix Bussi{\`{e}}res, Nicolas Sangouard, Mikael Afzelius, Hugues {De Riedmatten}, Christoph Simon, and Wolfgang Tittel.
\newblock {Prospective applications of optical quantum memories}.
\newblock {\em Journal of Modern Optics}, 60(18):1519--1537, 2013.

\bibitem{Ma2020}
Lijun Ma, Oliver Slattery, and Xiao Tang.
\newblock {Optical Quantum Memory and its Applications in Quantum Communication Systems}.
\newblock {\em Journal of research of the National Institute of Standards and Technology}, 125, 2020.

\bibitem{Hosseini2011}
M.~Hosseini, G.~Campbell, B.~M. Sparkes, P.~K. Lam, and B.~C. Buchler.
\newblock {Unconditional room-temperature quantum memory}.
\newblock {\em Nature Physics}, 7(10):794--798, 2011.

\bibitem{Finkelstein2018a}
Ran Finkelstein, Eilon Poem, Ohad Michel, Ohr Lahad, and Ofer Firstenberg.
\newblock {Fast, noise-free memory for photon synchronization at room temperature}.
\newblock {\em Science Advances}, 4(1):eaap8598, 2018.

\bibitem{Ebert2015}
M.~Ebert, M.~Kwon, T.~G. Walker, and M.~Saffman.
\newblock {Coherence and Rydberg Blockade of Atomic Ensemble Qubits}.
\newblock {\em Physical Review Letters}, 115(9):093601, 2015.

\bibitem{Jen2011}
H.~H. Jen and T.~A.~B. Kennedy.
\newblock {Efficiency of light-frequency conversion in an atomic ensemble}.
\newblock {\em Physical Review A}, 82(2), 2011.

\bibitem{Maring2018}
Nicolas Maring, Dario Lago-Rivera, Andreas Lenhard, Georg Heinze, and Hugues de~Riedmatten.
\newblock {Quantum frequency conversion of memory-compatible single photons from 606 nm to the telecom C-band}.
\newblock {\em Optica}, 5(5):507, 2018.

\bibitem{Zaske2012}
Sebastian Zaske, Andreas Lenhard, Christian~A. Ke{\ss}ler, Jan Kettler, Christian Hepp, Carsten Arend, Roland Albrecht, Wolfgang~Michael Schulz, Michael Jetter, Peter Michler, and Christoph Becher.
\newblock {Visible-to-telecom quantum frequency conversion of light from a single quantum emitter}.
\newblock {\em Physical Review Letters}, 109(14):147404, 2012.

\bibitem{Morrison2021}
Christopher~L. Morrison, Markus Rambach, Zhe~Xian Koong, Francesco Graffitti, Fiona Thorburn, Ajoy~K. Kar, Yong Ma, Suk~In Park, Jin~Dong Song, Nick~G. Stoltz, Dirk Bouwmeester, Alessandro Fedrizzi, and Brian~D. Gerardot.
\newblock {A bright source of telecom single photons based on quantum frequency conversion}.
\newblock {\em Applied Physics Letters}, 118(17):174003, 2021.

\bibitem{Becerra}
F~E Becerra, R~T Willis, S~L Rolston, and L~A Orozco.
\newblock {Nondegenerate four-wave mixing in rubidium vapor: The diamond configuration}.
\newblock {\em Physical Review A}, 78:013834, 2008.

\bibitem{Guo2016}
Xiang Guo, Chang~Ling Zou, Hojoong Jung, and Hong~X. Tang.
\newblock {On-Chip Strong Coupling and Efficient Frequency Conversion between Telecom and Visible Optical Modes}.
\newblock {\em Physical Review Letters}, 117(12):123902, 2016.

\bibitem{Hausmann2014}
B.~J.M. Hausmann, I.~Bulu, V.~Venkataraman, P.~Deotare, and M.~Loncar.
\newblock {Diamond nonlinear photonics}.
\newblock {\em Nature Photonics}, 8(5):369--374, 2014.

\bibitem{Li2016}
Qing Li, Marcelo Davan{\c{c}}o, and Kartik Srinivasan.
\newblock {Efficient and low-noise single-photon-level frequency conversion interfaces using silicon nanophotonics}.
\newblock {\em Nature Photonics}, 10(6):406--414, 2016.

\bibitem{Yu2016}
Ite~A. Yu, Ying-Cheng Chen, Bo-Han Wu, Gang Wang, Yong-Fang Chen, and Chin-Yuan Lee.
\newblock {High conversion efficiency in resonant four-wave mixing processes}.
\newblock {\em Optics Express}, 24(2), 2016.

\bibitem{Liu2017}
Zi~Yu Liu, Jian~Ting Xiao, Jia~Kang Lin, Jun~Jie Wu, Jz~Yuan Juo, Chin~Yao Cheng, and Yong~Fan Chen.
\newblock {High-efficiency backward four-wave mixing by quantum interference}.
\newblock {\em Scientific Reports}, 7(1):1--9, 2017.

\bibitem{Whiting2018}
D~J Whiting, Renju~S Mathew, J~Keaveney, C~S Adams, and I~G Hughes.
\newblock {Four-wave mixing in a non-degenerate four-level diamond configuration in the hyperfine Paschen-Back regime}.
\newblock {\em Journal of Modern Optics}, 65(6):713--722, 2018.

\bibitem{Willis2009a}
R.~T. Willis, F.~E. Becerra, L.~A. Orozco, and S.~L. Rolston.
\newblock {Four-wave mixing in the diamond configuration in an atomic vapor}.
\newblock {\em Physical Review A}, 79(3):033814, 2009.

\bibitem{Chaneliere2006}
T.~Chaneli{\`{e}}re, D.~N. Matsukevich, S.~D. Jenkins, T.~A~B Kennedy, M.~S. Chapman, and A.~Kuzmich.
\newblock {Quantum Telecommunication Based on Atomic Cascade Transitions}.
\newblock {\em Physical Review Letters}, 96(9):093604, 2006.

\bibitem{Franke-Arnold2010}
S.~Franke-Arnold, A.~Vernier, E.~Riis, and A.~S. Arnold.
\newblock {Enhanced frequency up-conversion in Rb vapor}.
\newblock {\em Optics Express}, 18(16):17020--17026, 2010.

\bibitem{Gogyan2010}
A.~Gogyan.
\newblock {Qubit transfer between photons at telecom and visible wavelengths in a slow-light atomic medium}.
\newblock {\em Physical Review A}, 81(2):024304, 2010.

\bibitem{Willis2011}
R~T Willis, F~E Becerra, L~A Orozco, and S~L Rolston.
\newblock {Photon statistics and polarization correlations at telecommunications wavelengths from a warm atomic ensemble}.
\newblock {\em Optics Express}, 19(15):14632, 2011.

\bibitem{Radnaev2010}
A.~G. Radnaev, Y.~O. Dudin, R.~Zhao, H.~H. Jen, S.~D. Jenkins, A.~Kuzmich, and T.~A.~B. Kennedy.
\newblock {A quantum memory with telecom-wavelength conversion}.
\newblock {\em Nature Physics}, 6(11):894, 2010.

\bibitem{Tseng2024}
Po~Han Tseng, Ling~Chun Chen, Jiun~Shiuan Shiu, and Yong~Fan Chen.
\newblock {Quantum interface for telecom frequency conversion based on diamond-type atomic ensembles}.
\newblock {\em Physical Review A}, 109(4):043716, 2024.

\bibitem{Boyd}
Robert~W. Boyd.
\newblock {\em Nonlinear optics}.
\newblock Academic Press, San Diego, fourth edition. edition, 2020.

\bibitem{Piotrowicz}
Micha{\l}~J Piotrowicz, Adam Black, and Mark Bashkansky.
\newblock Conversion from telecom to atomic photons by four-wave mixing in a warm rb cell.
\newblock In {\em Conference on Lasers and Electro-Optics}, page FW4C.4. Optica Publishing Group, 2020.

\bibitem{Kwolek2021}
Jonathan Kwolek, Adam Black, and Mark Bashkansky.
\newblock {Conversion from atomic to telecom photons by four-wave mixing in optically pumped warm Rb}.
\newblock {\em Frontiers in Optics and Laser Science}, page JTh5A.8, 2021.

\bibitem{ZhangFWM}
Wei-Hang Zhang, Ying-Hao Ye, Lei Zeng, Ming-Xin Dong, En-Ze Li, Jing-Yuan Peng, Yan Li, Dong-Sheng Ding, and Bao-Sen Shi.
\newblock {Telecom-wavelength conversion in a high optical depth cold atomic system}.
\newblock {\em Optics Express}, 31(5):8042, 2023.

\bibitem{Sparkes2013a}
B.~M. Sparkes, J.~Bernu, M.~Hosseini, J.~Geng, Q.~Glorieux, P.~A. Altin, P.~K. Lam, N.~P. Robins, and B.~C. Buchler.
\newblock {Gradient echo memory in an ultra-high optical depth cold atomic ensemble}.
\newblock {\em New Journal of Physics}, 15(8):085027, 2013.

\bibitem{Halfmann2014}
Thomas Halfmann, Thorsten Peters, and Frank Blatt.
\newblock {One-dimensional ultracold medium of extreme optical depth}.
\newblock {\em Optics Letters}, 39(3), 2014.

\bibitem{Poem2015}
Krzysztof~T. Kaczmarek, Dylan~J. Saunders, Michael~R. Sprague, W.~Steven Kolthammer, Amir Feizpour, Patrick~M. Ledingham, Benjamin Brecht, Eilon Poem, Ian~A. Walmsley, and Joshua Nunn.
\newblock {Ultrahigh and persistent optical depths of cesium in Kagom{\'{e}}-type hollow-core photonic crystal fibers}.
\newblock {\em Optics Letters}, 40(23), 2015.

\bibitem{Sprague2013}
Michael~R. Sprague, Duncan~G. England, Amir Abdolvand, Joshua Nunn, Xian~Min Jin, W.~{Steven Kolthammer}, Marco Barbieri, Bruno Rigal, Patrick~S. Michelberger, Tessa~F.M. Champion, Philip St~J. Russell, and Ian~A. Walmsley.
\newblock {Efficient optical pumping and high optical depth in a hollow-core photonic-crystal fibre for a broadband quantum memory}.
\newblock {\em New Journal of Physics}, 15(5):055013, 2013.

\bibitem{Perrella2018}
C.~Perrella, P.~S. Light, S.~Afshar Vahid, F.~Benabid, and A.~N. Luiten.
\newblock {Engineering Photon-Photon Interactions within Rubidium-Filled Waveguides}.
\newblock {\em Physical Review A}, 9(4):044001, 2018.

\bibitem{Donvalkar2014}
Prathamesh~S. Donvalkar, Vivek Venkataraman, St{\'{e}}phane Clemmen, Kasturi Saha, and Alexander~L. Gaeta.
\newblock {Frequency translation via four-wave mixing Bragg scattering in Rb filled photonic bandgap fibers}.
\newblock {\em Optics Letters}, 39(6), 2014.

\bibitem{Culver2016}
R.~Culver, A.~Lampis, B.~Megyeri, K.~Pahwa, L.~Mudarikwa, M.~Holynski, Ph~W. Courteille, and J.~Goldwin.
\newblock {Collective strong coupling of cold potassium atoms in a ring cavity}.
\newblock {\em New Journal of Physics}, 18(11):113043, 2016.

\bibitem{Rapol2002}
U.~D. Rapol and Vasant Natarajan.
\newblock {Precise measurement of hyperfine intervals using avoided crossing of dressed states}.
\newblock {\em Europhysics Letters}, 60(2):195, 2002.

\bibitem{Perrella2012}
C.~Perrella, P.~S. Light, T.~M. Stace, F.~Benabid, and A.~N. Luiten.
\newblock High-resolution optical spectroscopy in a hollow-core photonic crystal fiber.
\newblock {\em Physical Review A}, 85:012518, 2012.

\bibitem{Perrella2013}
C.~Perrella, P.~S. Light, J.~D. Anstie, T.~M. Stace, F.~Benabid, and A.~N. Luiten.
\newblock High-resolution two-photon spectroscopy of rubidium within a confined geometry.
\newblock {\em Physical Review A}, 87:013818, 2013.

\bibitem{SteckRb85}
D.~A. Steck.
\newblock {Rubidium 85 D Line Data}.
\newblock {\em http://steck.us/alkalidata (revision 2.1.5, 19 September 2012)}, 2012.

\bibitem{Duspayev2023}
A~Duspayev and G.~Raithel.
\newblock {Spectroscopy of the 85Rb 4D3/2 state for hyperfine-structure determination}.
\newblock {\em New Journal of Physics}, 25(9):093015, 2023.

\bibitem{DeSouza2022}
M~P~M {De Souza}, A~A~C {De Almeida}, and S~S Vianna.
\newblock {Dynamic Stark shift in Doppler-broadened four-wave mixing}.
\newblock {\em Physical Review A}, 105:53128, 2022.

\bibitem{Krehlik2022}
Tomasz Krehlik, Tomasz Krehlik, Artur Stabrawa, Rafa{\l} Gartman, Krzysztof~T. Kaczmarek, Robert L{\"{o}}w, and Adam Wojciechowski.
\newblock {Zeeman optical pumping of $^{87}$Rb atoms in a hollow-core photonic crystal fiber}.
\newblock {\em Optics Letters}, 47(21), 2022.

\bibitem{Akulshin2011}
A.~M. Akulshin, A.~A. Orel, and R.~J. McLean.
\newblock {Collimated blue light enhancement in velocity-selective pumped Rb vapour}.
\newblock {\em Journal of Physics B}, 45(1):015401, 2011.

\bibitem{Wen2014}
Feng Wen, Huaibin Zheng, Xinxin Xue, Haixia Chen, Jianping Song, and Yanpeng Zhang.
\newblock {Electromagnetically induced transparency-assisted four-wave mixing process in the diamond-type four-level atomic system}.
\newblock {\em Optical Materials}, 37(C):724--726, 2014.

\bibitem{Deng2002}
L.~Deng, M.~Kozuma, E.~W. Hagley, and M.~G. Payne.
\newblock {Opening Optical Four-Wave Mixing Channels with Giant Enhancement Using Ultraslow Pump Waves}.
\newblock {\em Physical Review Letters}, 88(14):143902, 2002.

\bibitem{Nasti2022}
Umberto Nasti, Hesham Sakr, Ian~A. Davidson, Francesco Poletti, and Ross~J. Donaldson.
\newblock {Utilizing broadband wavelength-division multiplexing capabilities of hollow-core fiber for quantum communications}.
\newblock {\em Applied Optics}, 61(30):8959, 2022.

\bibitem{Hosen:22}
Md.~Sarwar Hosen, Abdul Khaleque, Kumary Sumi~Rani Shaha, Lutfun~Nahar Asha, Azra~Sadia Sultana, Ruhana Nishad, and Md.~Tarek Rahman.
\newblock Highly birefringent polarization maintaining low-loss single-mode hollow-core antiresonant fiber.
\newblock {\em Opt. Continuum}, 1(10):2167--2184, 2022.

\bibitem{Fan2017}
Danyun Fan, Zhiqiang Jin, Guanghui Wang, Fei Xu, Yanqing Lu, Dora Juan~Juan Hu, Lei Wei, Ping Shum, and Xuping Zhang.
\newblock {Extremely High-Efficiency Coupling Method for Hollow-Core Photonic Crystal Fiber}.
\newblock {\em IEEE Photonics Journal}, 9(3):7104108, 2017.

\bibitem{Suslov2021}
Dmytro Suslov, Mat{\v{e}}j Komanec, Eric~R. {Numkam Fokoua}, Daniel Dousek, Ailing Zhong, Stanislav Zv{\'{a}}novec, Thomas~D. Bradley, Francesco Poletti, David~J. Richardson, and Radan Slav{\'{i}}k.
\newblock {Low loss and high performance interconnection between standard single-mode fiber and antiresonant hollow-core fiber}.
\newblock {\em Scientific Reports}, 11(1):1--9, 2021.

\bibitem{Sprague2014}
M.~R. Sprague, P.~S. Michelberger, T.~F.M. Champion, D.~G. England, J.~Nunn, X.~M. Jin, W.~S. Kolthammer, A.~Abdolvand, P.~St~J. Russell, and I.~A. Walmsley.
\newblock {Broadband single-photon-level memory in a hollow-core photonic crystal fibre}.
\newblock {\em Nature Photonics}, 8(4):287--291, 2014.

\bibitem{Londero2009}
P.~Londero, V.~Venkataraman, A.~R. Bhagwat, A.~D. Slepkov, and A.~L. Gaeta.
\newblock {Ultralow-power four-wave mixing with Rb in a hollow-core photonic band-gap fiber}.
\newblock {\em Physical Review Letters}, 103(4):043602, 2009.

\end{thebibliography}

\end{document}